\documentclass{aa} %[referee]
\usepackage[varg]{txfonts}
\usepackage{natbib}
\bibpunct{(}{)}{;}{a}{}{,} % to follow the A&A style
\usepackage{ifthen}
\usepackage{graphicx}
\usepackage[AA]{Math_Linc}
\usepackage{hyperref}
\hypersetup{colorlinks=true, citecolor=blue, linkcolor=blue, urlcolor=blue}

\defcitealias{Fan_etal_2010}{FSL10}
\newcommand{\for}[1]{Eq.~(\ref{#1})}
\newcommand{\fig}[1]{Fig.~\ref{#1}}
\newcommand{\figFull}[1]{Figure~\ref{#1}}
\newcommand{\sect}[1]{Sect.~\ref{#1}}

\title{A new model to predict weak-lensing peak counts}
\subtitle{I. Comparison with $N$-body simulations\thanks{The \textsc{Camelus} source code is released via the website \url{http://www.cosmostat.org/software/camelus/}}}
\titlerunning{A new model to predict weak-lensing peak counts I.}
\author{Chieh-An Lin \and Martin Kilbinger}
\authorrunning{C.-A. Lin \& M. Kilbinger}
\institute{
	Service d'Astrophysique, CEA Saclay, Orme des Merisiers, B\^at 709, 91191 Gif-sur-Yvette, France\\
	\email{\texttt{chieh-an.lin@cea.fr}}
} 
\date{Received 20 October 2014 / Accepted 20 January 2015}

\abstract%
	{Weak-lensing peak counts have been shown to be a powerful tool for cosmology. They provide non-Gaussian information of large scale structures and are complementary to second-order statistics.}
	{We propose a new flexible method for predicting weak-lensing peak counts, which can be adapted to realistic scenarios, such as a real source distribution, intrinsic galaxy alignment, mask effects, and photo-$z$ errors from surveys. The new model is also suitable for applying the tomography technique and nonlinear filters.}
	{A probabilistic approach to modeling peak counts is presented. First, we sample halos from a mass function. Second, we assign them density profiles. Third, we place those halos randomly on the field of view. The creation of these ``fast simulations'' requires much less computing time than do $N$-body runs. Then, we perform ray-tracing through these fast simulation boxes and select peaks from weak-lensing maps to predict peak number counts. The computation is achieved by our \textsc{Camelus} algorithm.}
	{We compare our results to $N$-body simulations to validate our model. We find that our approach is in good agreement with full $N$-body runs. We show that the lensing signal dominates shape noise and Poisson noise for peaks with S/N between 4 and 6. Also, counts from the same S/N range are sensitive to $\Omega_\mmmm$ and $\sigma_8$. We show how our model can distinguish between various combinations of those two parameters.}
	{In this paper, we offer a powerful tool for studying weak-lensing peaks. The potential of our forward model is its high flexibility, which makes the using peak counts under realistic survey conditions feasible.}
	

\begin{document}
\maketitle

\section{Introduction}
\label{sec:intro}

Weak gravitational lensing (WL) probes matter structures in the Universe. It contains information from the linear growth of structures to the recent highly nonlinear evolution, going from scales of hundreds of Mpc down to sub-Mpc levels. Until now, most studies have focused on two-point-correlation functions, but the non-Gaussianity of WL cannot be ignored if one aims for a deep understanding of cosmology.

One simple way to extract higher order WL information is peak counting. Peaks are defined as local maxima of the projected mass measurement. They are particularly interesting for at least two reasons. First, peaks are tracers of high-density regions. While other tracers of halo mass such as optical richness, X-ray luminosity or temperature, or the SZ Compton-$y$ parameter depend on scaling relations and often require assumptions about the dynamical state of galaxy clusters such as isothermal equilibrium and relaxedness, lensing does not. It therefore provides us with a direct way to study cosmology with the cluster mass function. Second, the lensing signal is highly non-Gaussian, and two-point-function-only studies deprive one of the information richness beyond second order. For example, \cite{Dietrich_Hartlap_2010} show that parameter constraints can be highly improved by joining peak counts and using second-order statistics, and \cite{Pires_etal_2012} find that peak counts capture more than the convergence skewness and kurtosis of the non-Gaussian information. Another advantage of WL peaks is information about the halo profile. \citet{Mainini_Romano_2014} showed that combining peak information with other cosmological probes provides an interesting way to study the mass-concentration relation.

\begin{figure*}[htb]
	\centering
	\includegraphics[width=16cm]{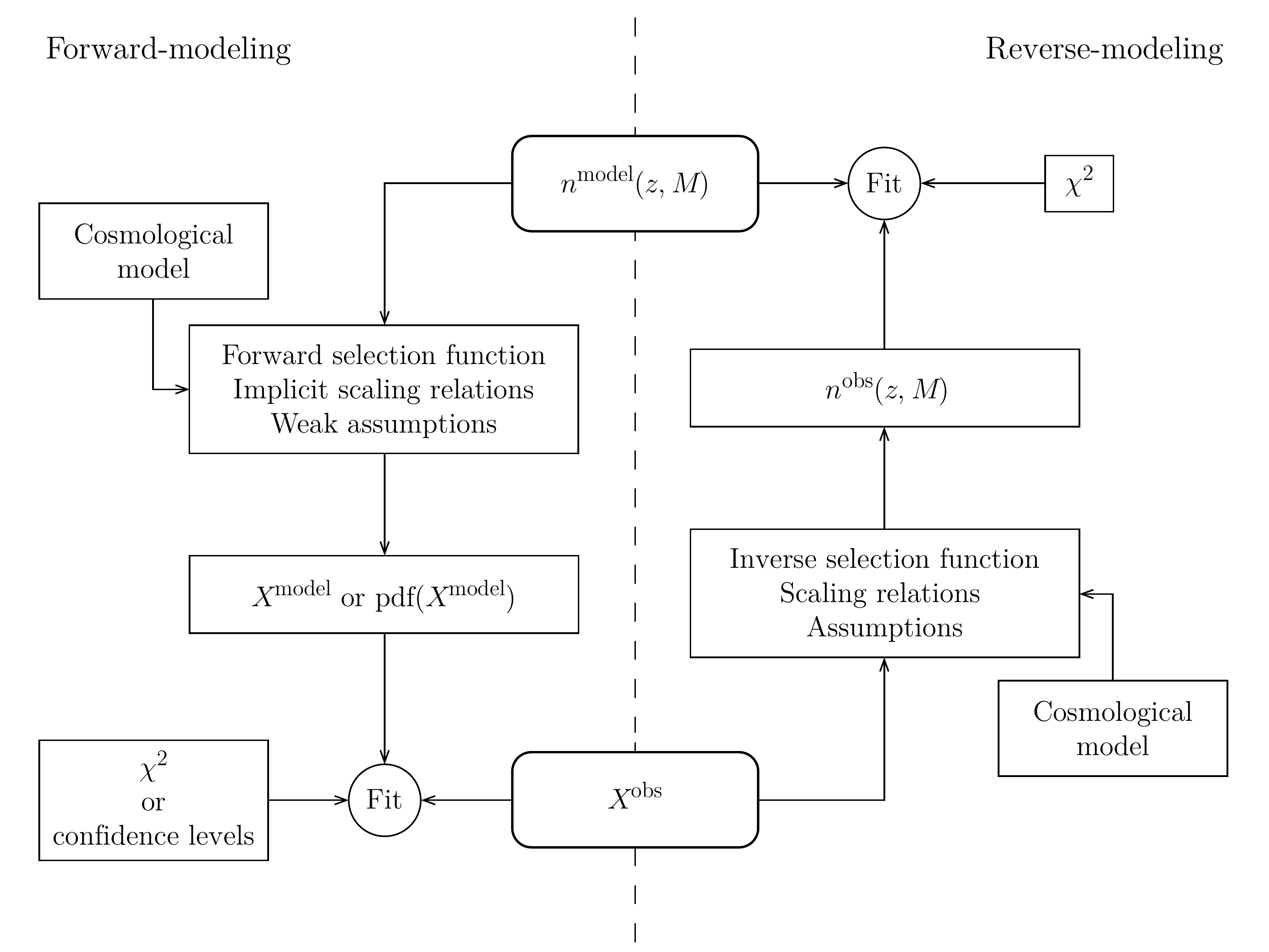}
	\caption{Forward- and reverse-modeling diagram for the mass function studies. Two different approaches to establishing links between the  theoretical mass function (the upper round rectangle) and the observables (the lower round rectangle) are to compare the observable mass function $n^\mathrm{obs}$ with the theoretical one (reverse modeling), or to compare ``predicted'' observable values $X^\mathrm{model}$ with observations (forward modeling). In this paper, the forward modeling is adopted and we propose a new method to ``predict'' peak conuts.}
	\label{fig:diagram}
\end{figure*}

For studies of the mass function via X-ray or the SZ effect, most works have adapted a reverse-fitting approach. This means that from diverse observables, one first establishes the observed mass function and then fit it with a theoretical model. To extract the mass function, this process needs to reverse the effect of selection functions, to use scaling relations, and to make further assumptions about sample properties. Alternatively, one can proceed with a forward-modeling approach: starting from an analytical mass function, we compute predicted values for observables and compare them to the data to carry out parameter fits (\fig{fig:diagram}). The corresponding forward application of selection functions is typically much simpler than its reverse. Moreover, instrumental effects can be easily included, and model uncertainties can be marginalized over. Forward modeling requires well-motivated models of physical phenomena, which is challenging in the case of observables derived from baryonic physics, yet \cite{Clerc_etal_2012} still provide a forward analysis from X-ray observations. For WL peak counts, however, computing the observable prediction is more straightforward, as long as using some appropriate assumptions.

One of the difficulties of predicting WL peak counts is that peaks can come from several mass overdensities at various redshifts due to projection effects \citep{Jain_VanWaerbeke_2000, Hennawi_Spergel_2005, Kratochvil_etal_2010}. This makes counting nonadditive even in the linear regime, and the prediction becomes less trivial. To overcome this ambiguous effect, some previous works have used $N$-body simulations, e.g., \cite{Dietrich_Hartlap_2010}. They perform peak counts from $N$-body runs with different paremeter sets to obtain confidence contours for constraints. However, since $N$-body simulations are very costly in terms of computation time, input parameter sets should be carefully chosen, and an interpolation of results is needed. Thus the resolution in the parameter space is limited, and the Fisher matrix is only available for the fiducial parameters.

Alternatively, there have been several attempts at peak-count modeling. \cite{Maturi_etal_2010} propose to study contiguous areas of high-signal regions instead of peaks, and provide a model that predicts the amount of this alternative observable. Meanwhile, \citet*[][hereafter \citetalias{Fan_etal_2010}]{Fan_etal_2010} propose a model for convergence peaks by supposing at most one halo exists on each line of sight. Both models are analytical and based on calculation from Gaussian random field theory. A comparison of the \citetalias{Fan_etal_2010} model with observation has been shown by \cite{Shan_etal_2014}, using the data from the Canada-France-Hawaii Telescope Stripe 82 Survey.

However, these models encounter difficulties for additional complications and subtleties. On one hand both models require Gaussian noise and linear filters, otherwise the Gaussian random field theory becomes invalid. As a result, non-linear, optimized reconstruction methods of the projected overdensity are automatically excluded. On the other hand realistic scenarios, such as mask effects and intrinsic ellipticity alignment, introduce asymmetric changes into the peak counts. The impact of these additional effects are unpredictable in purely analytical models. This encourages us to propose a new model for WL peak counts.

In this paper, we adopt a probabilistic approach to forecasting peak counts. This can be handled by our \textsc{Camelus} algorithm (Counts of Amplified Mass Elevations from Lensing with Ultrafast Simulation). Unlike $N$-body simulations which are very time-consuming, we create ``fast simulations'' by sampling halos from the mass function. The only requirement is a cosmology with a known mass function and halo mass profiles. To validate this method and to justify various hypotheses that our model makes, we compare results from our fast simulations to those from $N$-body runs. This approach is similar to the sGL model of \cite{Kainulainen_Marra_2009, Kainulainen_Marra_2011, Kainulainen_Marra_2011a}, where they show that the stochastic process provides a quick and accurate way to recover the lensing signal distribution.

The outline of this paper is as follows. In \sect{sec:theory}, we recall some of the WL formalism and theoretical support for our model. In \sect{sec:model}, a full description of our model is given. In \sect{sec:simu}, we give the details concerning the $N$-body and the ray-tracing simulations. Finally, the results are presented in \sect{sec:results}, before we summarize and conclude in \sect{sec:conclu}.

\section{Theoretical basics}
\label{sec:theory}

In this section, we define the formalism necessary for our analysis. To model the convergence field lensed by halos, we need to specify their profile, projected mass, and distribution in mass and redshift, which is the mass function.

\subsection{Weak lensing convergence}
\label{subsec:WL}

Observationally, galaxy shape distortions can be displayed at linear order in the form of the lensing distortion matrix $\mathcal{A}$. For an angular position $\btheta$, $\mathcal{A}(\btheta)$ is given by
\begin{align}
	\mathcal{A}(\btheta) =
	\begin{pmatrix}
		1-\kappa - \gamma_1 &          - \gamma_2\\
		         - \gamma_2 & 1-\kappa + \gamma_1
	\end{pmatrix},
\end{align}
which defines two WL observables: convergence $\kappa$ and shear $\gamma$. The latter is a complex number given by $\gamma = \gamma_1 + \iiii \gamma_2$. This linearization of the light distortion can be calculated explicitly in general relativity. Accordingly, the matrix elements are linked to second derivatives of the Newtonian gravitational potential $\phi$ by
\begin{align}
	\mathcal{A}_{ij}(\btheta) = \delta_{ij} - \frac{2}{\cccc^2}\int_0^w \dddd w'\ \frac{f_K(w-w')f_K(w')}{f_K(w)}\ \phi_{,ij}\big(f_K(w')\btheta, w'\big),
\end{align}
where $f_K$ is the comoving transverse distance and $\delta_{ij}$ the Kronecker delta. In particular, an explicit expression of $\kappa$ is given as follows \citep[see, e.g.,][]{Schneider_etal_1998},
\begin{align}\label{for:WL_4}
	\kappa(\btheta, w) = \frac{3H^2_0 \Omega_\mmmm}{2\cccc^2} \int_0^w \dddd w'\ \frac{f_K(w-w')f_K(w')}{f_K(w)} \frac{\delta\big( f_K(w')\btheta, w' \big)}{a(w')},
\end{align}
where $H_0$ is the Hubble parameter, $\Omega_\mmmm$ the matter density, $\cccc$ the speed of light, $a(w')$ represents the scale factor at the epoch to which the comoving distance from now is $w'$, and $\delta$ is the matter density contrast.

\subsection{Halo density profile and its projected mass}
\label{subsec:NFW}

Consider now a dark matter (DM) halo with a Navarro-Frenk-White (NFW) density profile \citep{Navarro_etal_1996, Navarro_etal_1997}, given by
\begin{align}\label{for:WL_5}
	\rho(r) = \frac{\rho_\ssss}{(r/r_\ssss)^\alpha (1+r/r_\ssss)^{3-\alpha}},
\end{align}
where $\rho_\ssss$ and $r_\ssss$ are the characteristic mass density and the scale radius of the halo, respectively, and $\alpha$ is the inner slope parameter. The concentration parameter $\cNFW$ is defined as the ratio of the virial radius to the scale radius, $\cNFW = r_\vir/r_\ssss$. We assume the following expression \citep[proposed by][]{Takada_Jain_2002}:
\begin{align}
	\cNFW(z,M) = \frac{c_0}{1+z}\left(\frac{M}{M_\star}\right)^{-\beta},
\end{align}
where $M$ is the halo mass and $M_\star$ the pivot mass such that $\delta_\cccc(z=0) = \sigma(M_\star)$, with $\delta_\cccc$ the threshold overdensity for the spherical collapse model, and $\sigma^2(M)$ is the variance of the density contrast fluctuation smoothed with a top-hat sphere with radius $R$ such that $M=\bar{\rho}_0(4\pi/3)R^3$. 

In this paper, we take $c_0 = 8$, $\alpha = 1$, and $\beta = 0.13$. The value of $\alpha$ corresponds to the classical NFW profile. The value of $\beta$ is provided by \cite{Bullock_etal_2001}, and $c_0$ corresponds to the best-fit value, using $r_\vir$, $r_\ssss$, $z$, $M$ derived from the $N$-body simulations that we use and fixing $\beta$. For $\delta_\cccc$, we use the fitting formula of \citet{Weinberg_Kamionkowski_2003} with
\begin{align}
	\delta_\cccc(z) = \frac{3(12\pi)^{2/3}}{20} \left(1+\alpha\log_{10}\Omega_\mmmm(z)\right),
\end{align}
and 
\begin{align}
	\alpha = 0.353w^4 + 1.044w^3 + 1.128w^2 + 0.555w + 0.131.
\end{align}

Lensing by an NFW halo is characterized by its projected mass. More precisely, defining the scale angle $\theta_\ssss = r_\ssss/\Dl$ as the ratio of the scale radius to the angular diameter distance $\Dl$ between lens and observer, we get \citep[following][]{Bartelmann_1996, Takada_Jain_2003a} \footnote{The convention of \cite{Takada_Jain_2003a} is different from ours. Their $d_\AAAA$ is actually $f_K$ in our notation, and they also express the virial radius $r_\vir$ in comoving coordinates.} \footnote{For computational purpose, $2\rho_\ssss r_\ssss = (Mf\cNFW^2) / (2\pi r_\vir^2)$, where $f = [\ln(1+\cNFW) - \cNFW/(1+\cNFW)]\inv$.}
\begin{align}\label{for:WL_1}
	\kappa_\proj(\btheta) = \frac{2\rho_\ssss r_\ssss}{\Sigma_\cccc} G\left(\frac{\theta}{\theta_\ssss}\right),
\end{align}
with
\begin{align}\label{for:WL_2}
	\Sigma_\cccc = \frac{\cccc^2}{4\pi\GGGG} \frac{\Ds}{\Dl\Dls},
\end{align}
where the quantities $\Ds$ and $\Dls$ are the angular diameter distances between source and observer, and lens and source, respectively, and 
\begin{align}\label{for:WL_3}
	G(x) =\left\{
	\begin{array}{l}
		\hspace{-0.5em}\displaystyle -\frac{1}{1-x^2}\frac{\scalebox{0.8}{$\sqrt{\cNFW^2-x^2}$}}{\cNFW+1} + \frac{1}{(1-x^2)^{3/2}}
		\arcosh\left[ \frac{x^2+\cNFW}{\scalebox{0.9}{$x(\cNFW+1)$}} \right]\\[3ex]
		\hfill \text{if $x<1$;}\\
		\displaystyle\frac{\scalebox{0.8}{$\sqrt{\cNFW^2-1}$}}{\cNFW+1} \cdot \frac{\cNFW+2}{3(\cNFW+1)} \hfill \text{if $x=1$;}\\[3ex]
		\displaystyle\frac{1}{x^2-1}\frac{\scalebox{0.8}{$\sqrt{\cNFW^2-x^2}$}}{\cNFW+1} - \frac{1}{(x^2-1)^{3/2}}\arccos\left[ \frac{x^2+\cNFW}{\scalebox{0.9}{$x(\cNFW+1)$}} \right]\\[3ex]
		\hfill \text{if $1<x\leq\cNFW$;}\\[1ex]
		\displaystyle 0 \hfill \text{if $x > \cNFW$.}
	\end{array}\right.
\end{align}
We have truncated the projected mass distribution at $\theta = \cNFW\theta_\ssss$. Equation (\ref{for:WL_1}) is used and computed for the ray-tracing simulations with NFW halos.

\subsection{Halo mass function}
\label{subsec:massFct}

The halo mass function $n(z,\lessM)$ indicates the halo number density with mass less than $M$ at redshift $z$ \footnote{Some papers define the mass function as $\tilde{n}(z,M)$, where $\tilde{n}(z,M) = \dddd n(z,\lessM) / \dddd M$.}, often characterized by a function $f(\sigma, z)$ as
\begin{align}
	f(\sigma, z) \equiv \frac{M}{\bar{\rho}_0}\frac{\dddd n(z,\lessM)}{\dddd\ln\sigma\inv(z,M)},
\end{align}
where $\bar{\rho}_0$ is the current matter density, and $\sigma(z,M)$ is defined as $\sigma(M)$ multiplied by the growth factor $D(z)$. In this study, we adopt the model proposed by \cite{Jenkins_etal_2001} in which a fit for $f$ is given as
\begin{align}
	f(\sigma) = 0.315 \exp\left[ -\left|\ln\sigma\inv + 0.61\right|^{3.8} \right].
\end{align}

\section{A new model for WL peak counts}
\label{sec:model}

\subsection{Probabilistic approach: fast simulations}
\label{subsec:ours}

Our model is based on the idea that we can replace $N$-body simulations with an alternative random process, such that the relevant observables are preserved, but the computation time is drastically reduced. We call this alternative process ``fast simulations'', which are produced by the following steps:
\begin{enumerate}
	\item generate halo masses by sampling from a mass function,
	\item assign density profiles to the halos,
	\item place the halos randomly on the field of view,
	\item perform ray-tracing simulation.
\end{enumerate}

One can notice that we have made two major hypotheses. First, we assume that diffuse, unbound matter, for example cosmological filaments, does not significantly contribute to peak counts. Second, we suppose that the spatial correlation of halos has a minor influence, since this correlation is broken down in fast simulations. Previous work has shown that correlated structures influence number and height of peaks by only a few percentage points \citep{Marian_etal_2010}. Furthermore, assuming a stochastical distribution of halos can lead to accurate predictions of the convergence probability distirbution function \citep{Kainulainen_Marra_2009}. One may also notice that halos can overlap in 3D space, and indeed we do not exclude this possibility. We test and validate these hypotheses in \sect{sec:results}, and discuss possible improvements to our model in \sect{sec:conclu}

Although we have chosen NFW profiles for the density of DM halos, using any halo profile model for which the projected mass is known is of course possible, such as triaxial halos or profiles offered by baryonic feedback \citep{Yang_etal_2013}. In addition, our prediction model is completely independent of the method by which peaks are extracted from the weak-lensing data. The same analysis can be applied to data (or $N$-body simulations + ray-tracing) and to fast simulations. Moreover, survey characteristics, such as masks, photometric redshift errors, PSF residuals, and other systematics, can be incorporated and forward-propagated as model uncertainties. Furthermore, the halo sampling technique is much faster than a full $N$-body run. For instance, it only takes a dozen seconds on a single-CPU desktop computer to generate a box that is large enough for our use (see specifications in \sect{subsec:fastSimu}).

This is a probabilistic approach to forecast peak counts, and we compare the convergence peaks obtained with those from full $N$-body runs in order to validate our forward model. This is described in \sect{subsec:validation}.

\subsection{Peak selection}
\label{subsec:peak}

In this paper, we focus on convergence peaks. We have followed a classical analysis used in former studies \citep[e.g.,][]{Hamana_etal_2004, Wang_etal_2009, Fan_etal_2010, Yang_etal_2011} to extract peaks.

First, we should highlight that $\kappa$ and $\kappa_\proj$ (respectively given by Eqs. \eqref{for:WL_4} and \eqref{for:WL_1}) do not follow the same definition. Actually, \for{for:WL_1} can be recovered by replacing $\delta$ with $\rho/\bar{\rho}$ in \for{for:WL_4}. This means that $\kappa_\proj$ does not take lensing by underdense regions into account and is shifted by a constant value, which corresponds to the mass-sheet degenerency. To obtain a model that is consistant with a zero-mean convergence field, we subtract the mean value of $\kappa_\proj$ from our convergence maps, so that
\begin{align}
	\kappa(\btheta) = \kappa_\proj(\btheta) - \overline{\kappa_\proj}.
\end{align}
We use this approximation throughout this study when ray-tracing is done with projected mass.

Consider now a reconstructed convergence field $\kappa_n(\btheta)$ in the absence of intrinsic ellipticity alignment. The presence of galaxy shape noise leads to the true lensing field $\kappa(\btheta)$ being contaminated by a linear additive noise field $n(\btheta)$, such that
\begin{align}\label{for:model_1}
	\kappa_n(\btheta) = \kappa(\btheta) + n(\btheta).
\end{align}
In general, $\kappa$ is dominated by $n$, and one way to suppress the noise is to apply a smoothing:
\begin{align}
	K_N(\btheta) \equiv (\kappa_n\ast W)(\btheta) = \int\dddd\btheta'\ \kappa_n(\btheta-\btheta')W(\btheta')
\end{align}
where $W(\btheta)$ is a window function, chosen to be Gaussian in this study as
\begin{align}
	W(\btheta) = \frac{1}{\pi\theta_\GGGG}\exp\left( -\frac{\theta^2}{\theta_\GGGG^2} \right),
\end{align}
which is specified by the smoothing scale $\theta_\GGGG$. We denote $K_N(\btheta)$, $K(\btheta)$, and $N(\btheta)$ as corresponding smoothed fields to \for{for:model_1}, such that
\begin{align}
	K_N(\btheta) = K(\btheta) + N(\btheta),
\end{align}
and set $\theta_\GGGG = 1$~arcmin in the following.

If intrinsic ellipticities are uncorrelated between source galaxies, $N(\btheta)$ can be described as a Gaussian random field \citep{Bardeen_etal_1986, Bond_Efstathiou_1987} for which the variance is related to the number of galaxies contained in the filter. This is given by \cite{VanWaerbeke_2000} as
\begin{align}\label{for:model_2}
	\sigma_\noise^2 = \frac{\sigma_\epsilon^2}{2}\frac{1}{2\pi n_\gggg \theta_\GGGG^2}.
\end{align}
Here, $n_\gggg$ is the source galaxy number density, and $\sigma_\epsilon^2 = \langle\epsilon_1^2\rangle + \langle\epsilon_2^2\rangle$ is the variance of the intrinsic ellipticity distribution. We then define the lensing S/N as
\begin{align}
	\nu(\btheta) \equiv \frac{K_N(\btheta)}{\sigma_\noise},
\end{align}
and the peaks are extracted from the $\nu$ field, defined as pixels that have a S/N value higher than their eight neighbors. This implies that peak analyses require S/N values on a well-defined grid (e.g., HEALPix grid). Furthermore, we suppose that source galaxies are uniformly distributed in this study, so $\sigma_\noise$ is a constant. However, this does not have to be true in general.

In summary, convergence peaks are selected by the following steps:
\begin{enumerate}
	\item compute the projected mass $\kappa_\proj(\btheta)$ by ray-tracing,
	\item subtract the mean to obtain $\kappa(\btheta)$,
	\item add the noise to obtain $\kappa_n(\btheta)$,
	\item smooth the field and acquire $K_N(\btheta)$,
	\item determine the S/N $\nu(\btheta)$, and
	\item select local maxima and compute the density $n_\peak(\nu)$.
\end{enumerate}
Only positive peaks are selected, and the analysis is based on the abundance histograms from peak counts. From fast simulation, through ray-tracing, to peak selection, the calculation is carried out by our \textsc{Camelus} algorithm.

\section{Simulations}
\label{sec:simu}

\subsection{$N$-body simulations}
\label{subsec:aardvark}

As provided by A. Evrard, the $N$-body simulations ``Aardvark'' have been used in this study. They were generated by \textsc{LGadget-2}, a DM-only version of \textsc{Gadget-2} \citep{Springel_2005}. The Aardvark parameters had been chosen to represent a WMAP-like $\Lambda$CDM cosmology, with $\Omega_\mathrm{m} = 0.23$, $\Omega_\Lambda = 0.77$, $\Omega_\mathrm{b} = 0.047$, $\sigma_8 = 0.83$, $h = 0.73$, $n_\mathrm{s} = 1.0$, and $w_0 = -1.0$.

The DM halos in Aardvark were identified using the \textsc{Rockstar} friends-of-friends code \citep{Behroozi_etal_2013}. The field of view is 859 deg$^2$. This corresponds to a HEALPix patch with $n_\mathrm{side} = 2$ \citep[for HEALPix, see][]{Gorski_etal_2005}.

\begin{figure*}[htb]
	\centering
	\includegraphics[width=17cm]{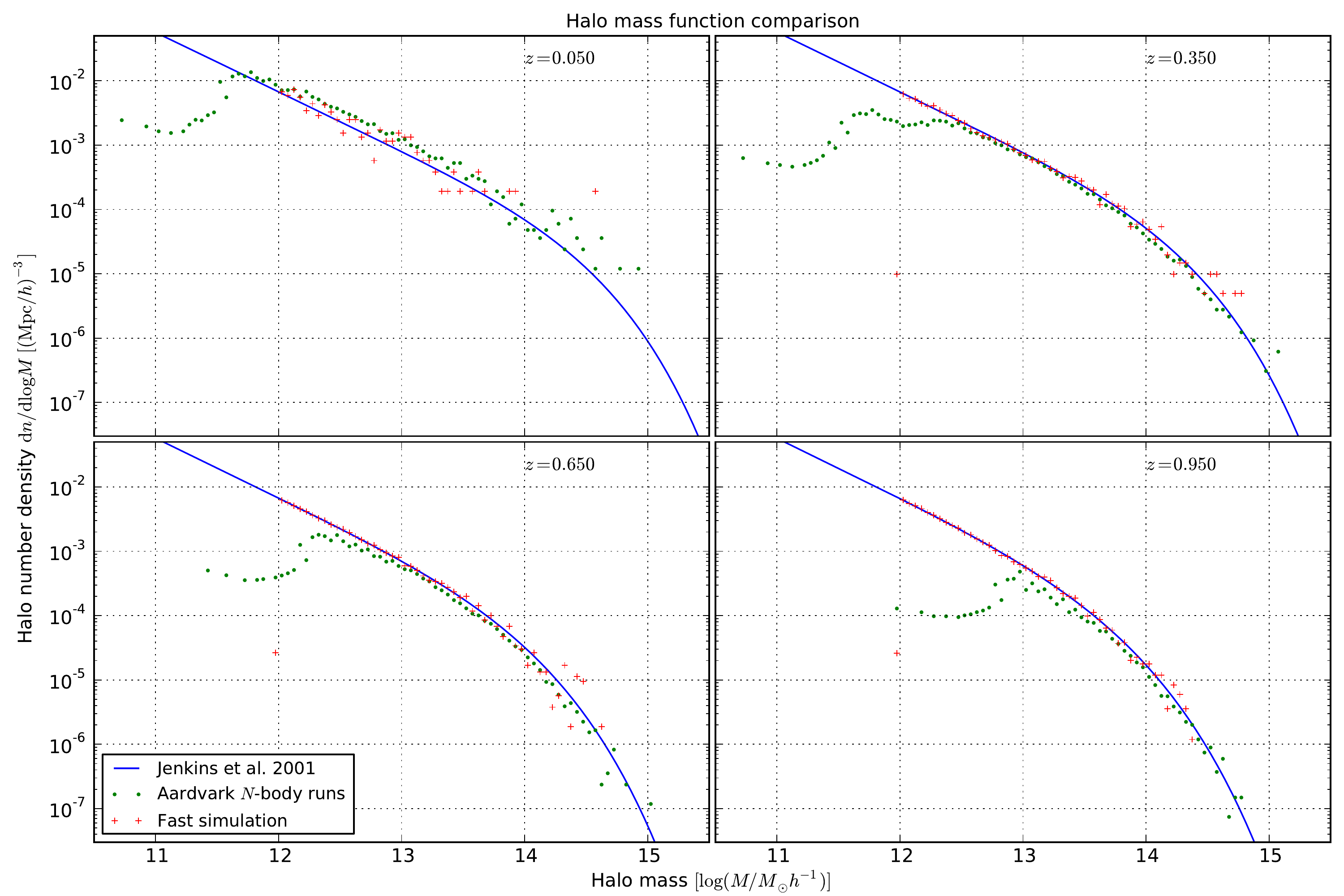}
	\caption{Comparison between an analytical mass function (blue line), halo mass histograms for $N$-body runs (green points), and sample histograms for realizing of the fast simulation (red crosses). The plots are drawn at 4 different redshift planes, and for each the thickness is $\dddd z=0.1$.}
	\label{fig:massComp}
\end{figure*}

\subsection{Fast simulations}
\label{subsec:fastSimu}

As described in Sect \ref{subsec:ours}, our model requires a mass function as input. We chose the model of \citet[][see \sect{subsec:massFct}]{Jenkins_etal_2001} to sample halos. This is done in ten redshift bins from $z =$~0 to 1. We set the sample mass range to the interval $\dix{12}$ and $\dix{17} M_\odot/h$. For each halo, the NFW parameters were set to be $(c_0, \alpha, \beta) = (8.0, 1.0, 0.13)$. We suggest seeing \sect{subsec:NFW} for their definitions.

\figFull{fig:massComp} shows an example of our halo samples, compared to the original mass function, and mass histograms established from the Aardvark simulations. Although halos with high mass can be $\dix{3}$--$\dix{5}$ times less populated than low-mass halos, our sampling is still in a perfect agreement with the original mass function. One may notice a shift and a tilt in the Aardvark halo mass function for low and high redshifts, however, in these regimes, the lensing efficiency is low because of the distance weight term $\Dl\Dls/\Ds$, so this mismatch is not very large.

\subsection{Ray-tracing simulations}
\label{subsec:RT}

For the Aardvark simulations, ray-tracing was performed with CALCLENS \citep{Becker_2013}. Galaxies were generated using ADDGALS (by M. Busha and R. Wechsler \footnote{\url{http://bitbucket.org/mbusha/addgals}}). Ray-tracing information is available only on a subset of 53.7 deg$^2$ (a HEALPix patch with $n_\mathrm{side}=8$), which is 16 times smaller than the halo field. In this study, only galaxies at redshift between 0.9 and 1.1 were chosen for drawing the convergence map. It led to an irregular map, and in order to clearly define eight neighbors to identify peaks, we used a 2D-linear interpolation to obtain $\kappa$ values on a grid. This was done after carrying out a projection to Cartesian coordinates.

For computational purposes, in order not to handle too many galaxies at a time, we split the field into four ``ray-tracing patches'', the size of which is 13.4 deg$^2$ each (corresponding to $n_\mathrm{side}=16$). We then project the coordinates with regard to the center of each patch using the Gnomonic projection. The size lengths of the ray-tracing patches are between 3.5 and 6.2~deg, so small enough to retain a good approximation.

For the fast simulations and the two intermediate cases that we study in \sect{subsec:validation}, source galaxies have a fixed redshift $z_\ssss=1.0$. They are regularly distributed on a HEALPix grid and placed at the center of pixels. Each ray-tracing pixel is a HEALPix patch with $n_\mathrm{side} =$ 16,384, for which the characteristic size is $\theta_\pix \approx$~0.215 arcmin. Thus, the galaxy number density is $n_\gggg=1/\theta_\pix^2=21.7$ arcmin$\invSq$. Ray-tracing for fast simulations is carried out after splitting and projection to Cartesian coordinates. There are 64 ray-tracing patches in a halo field, and each patch contains 1024 $\times$~1024 pixels. The convergence was computed using Eqs. \eqref{for:WL_1}, \eqref{for:WL_2}, and \eqref{for:WL_3}. As a remark, no mask was applied in this study.

\subsection{Adding noise}
\label{subsec:noise}

Shape noise $n(\btheta)$ is added to each pixel after we obtain $\kappa(\btheta)$ from $N$-body runs or fast simulations. It is modeled as a Gaussian random field with a top-hat filter with a size that corresponds to the pixel area $A_\pix$. The variance of this is given by \cite{VanWaerbeke_2000} as
\begin{align}
	\sigma_\pix^2 = \frac{\sigma_\epsilon^2}{2}\frac{1}{n_\gggg A_\pix}.
\end{align}

\begin{figure*}[htb]
	\centering
	\includegraphics[width=17cm]{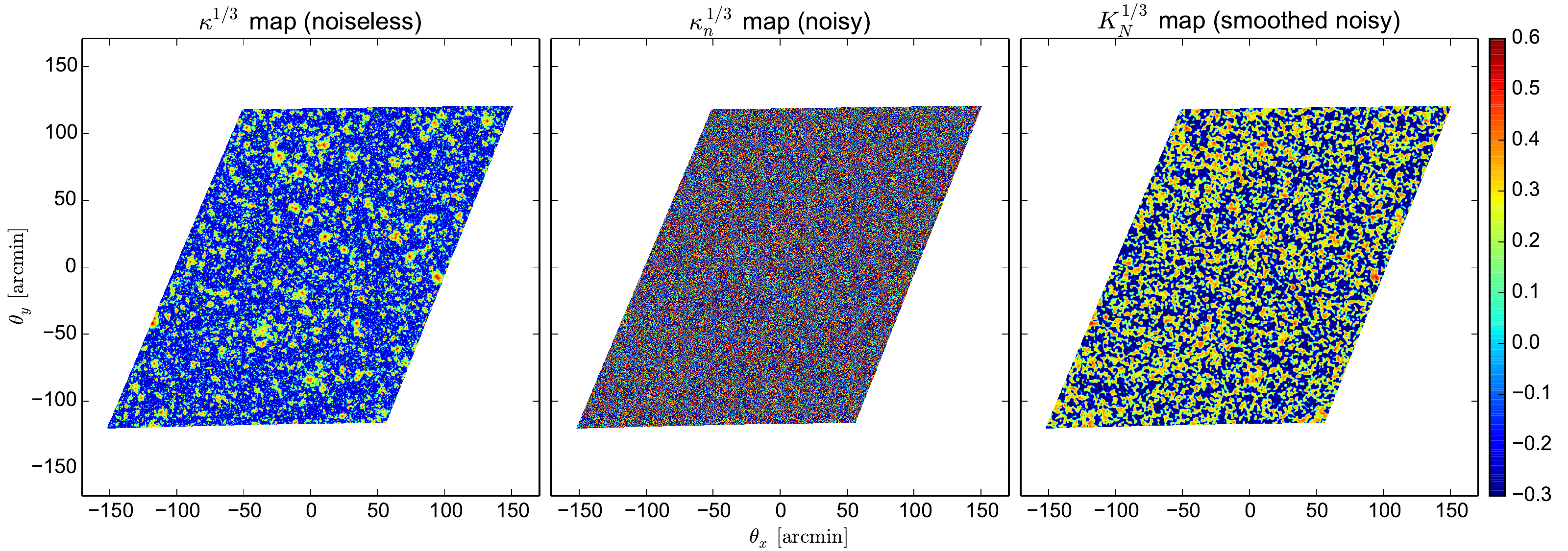}
	\caption{Patch of a map projected with regard to its center, taken from a realization of fast simulations. The left, middle, and right panels give the fields $\kappa^{1/3}$, $\kappa_n^{1/3}$, and $K_N^{1/3}$, respectively. We have taken the cubic root to emphasize the contrast. It is clear that the signal is completely dominated by the noise. Even though the smoothed map is quite different from the original one, the structures, high-signal regions are still conserved and traced.}
	\label{fig:HPMap}
\end{figure*}

We choose $\sigma_\epsilon=0.4$ which corresponds to a CFHTLenS-like survey, and $n_\gggg A_\pix$ is chosen to be 1 so that each pixel represents one galaxy. This leads to $\sigma_\pix\approx0.283$. We can also estimate $\sigma_\noise$ with \for{for:model_2} and obtain $\sigma_\noise \approx 0.024$. This shows that a real map is in general dominated by the noise (\fig{fig:HPMap}). Even for a peak at $\nu=5$, the lensing signal is only on the order of $\kappa=0.12$, less than half of the pixel noise amplitude.

\section{Results}
\label{sec:results}

\subsection{Validation of our model: comparison to $N$-body runs}
\label{subsec:validation}

To validate our model, we compare it to the $N$-body simulations. We compute peak abundance histograms from both simulations, together with two intermediate steps. This results in four cases in total:
\addtolength{\leftmargini}{2em}
\begin{itemize}
	\item[Case 1:] full $N$-body runs;
	\item[Case 2:] replacing $N$-body halos with NFW profiles with the same masses;
	\item[Case 3:] randomizing angular positions of halos from Case 2;
	\item[Case 4:] fast simulations, corresponding to our model.
\end{itemize}
\addtolength{\leftmargini}{-2em}

These cases form a progressive transition from full $N$-body runs toward our model. More precisely, Case 2 tests the hypothesis corresponding to the second step of our model (see \sect{subsec:ours}); i.e., diffuse, unbound matter contributes little to peak counts. Case 3 additionally tests the assumption made in the third step. (Halo clustering plays a minor role.) Finally, Case 4 completes our model with the missing first step. As a result, the halo population and their redshifts are identical to $N$-body runs in Cases 2 and 3.

\begin{figure*}[htb]
	\centering
	\includegraphics[width=17cm]{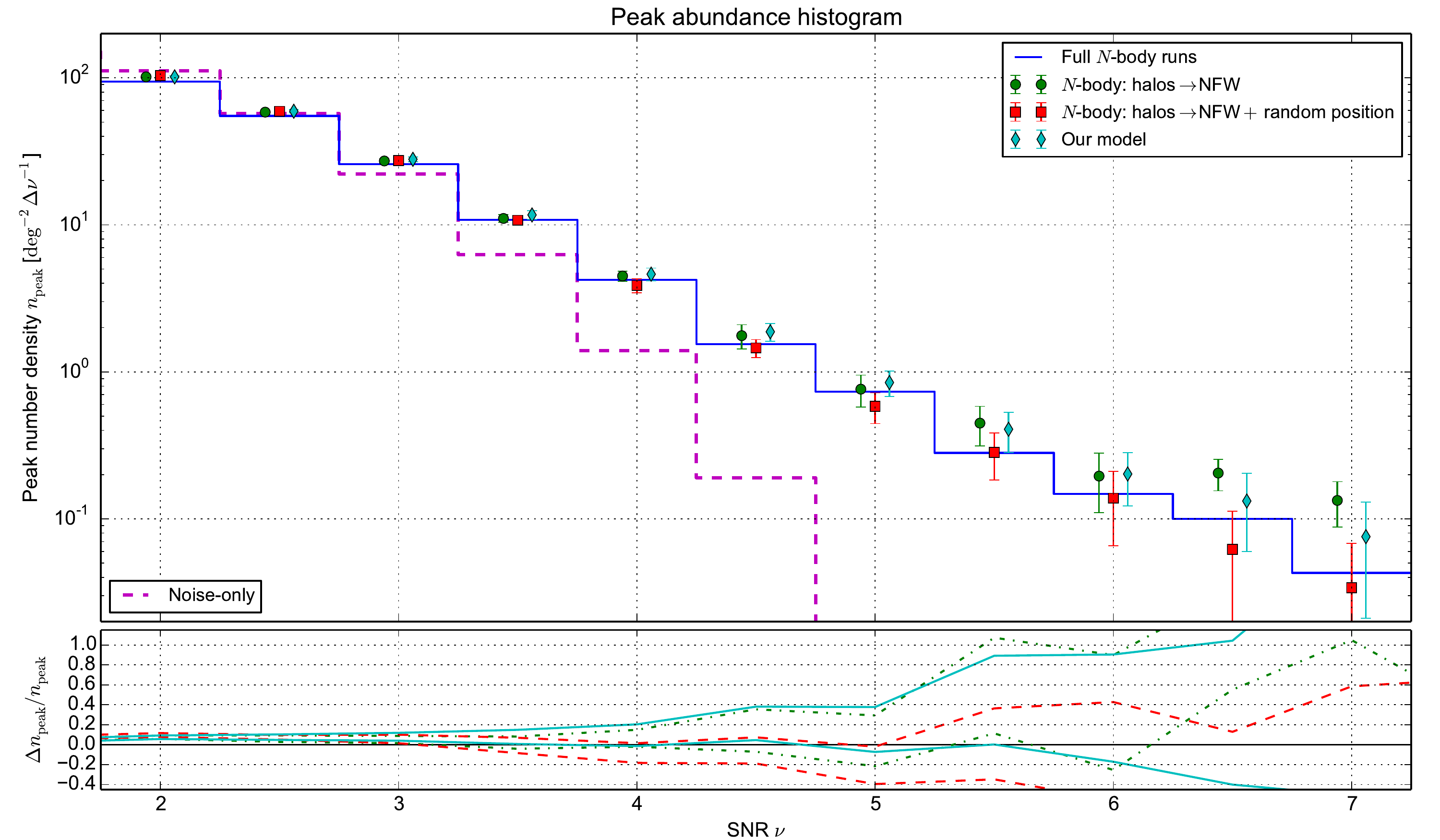}
	\caption{Comparison of the peak abundance from different cases. In the upper panel, blue solid line: full $N$-body runs (Case 1); green circles: replacement of halos by NFW profiles (Case 2); red squares: replacement of halos by NFW profiles and randomization of halo angular positions (Case 3); cyan diamonds: fast simulations, corresponding to our model (Case 4); magenta dashed line: peaks from noise-only maps. In the lower panel, we draw the upper and lower limits of error bars shifted with regard to the $N$-body values. This refers to the standard deviation over 4 maps (green dash-dotted line for Case 2) or 16 maps (red dashed line for Case 3, cyan solid line for Case 4). The field of view is 53.7 deg$^2$.}
	\label{fig:small_field}
\end{figure*}

\figFull{fig:small_field} shows the peak abundance histograms for all four cases. In this section, the field of view is 53.7 deg$^2$, since we are limited by the available information of ray-tracing for the $N$-body runs. For Cases 1 and 2, we compute the average in each histogram bin for eight noise maps. For Cases 3 and 4, this is done with eight realizations (of randomization and of fast simulations, respectively) and eight noise maps, thus 64 maps in total. The error bars therefore refer to the combination of the statistical fluctuation due to the random process and the shape noise uncertainty.

For low peaks with $\nu\leq 3.75$, we observe that $n_\peak(\nu)$ remains almost unchanged between the different cases. This is not suprising because in this regime, $n_\peak(\nu)$ is mainly contributed by noise. This argument is supported by the noise-only peak histogram. The lower panel of \fig{fig:small_field} shows that there exist some systematic overcounts in this regime on the order of 10\%. The cause of this bias is ambiguous. One possibility might be the use of NFW profiles for ray-tracing simulations. It might also come from the subtraction of the mean $\kappa$ value from the maps. We leave this to future studies. Another observation in this regime is that by adding the signal to the noise field, the number of peaks with $\nu\leq2.75$ decreases. This proves that the effect of noise is not additive for peak counts.

In the regime of $\nu\geq3.75$, we observe that replacement by NFW profiles enhances the peak counts, while position randomization introduces an opposite effect of a similar order of magnitude. The enhancement from Case 2 may be explained by the halo triaxiality. A spherical profile such as the NFW model may lead to an overestimation of the projected mass at the center of halos if the major axis is not aligned with the line of sight, and this would probably be the case for most of the $N$-body halos. It could also be an effect of the $M$-$c$ relation: we might overestimate $\cNFW$ for large $M$.

Comparing Cases 2 and 3, we discover that position randomization decreases peak counts by 10\% to 50\%. Apparently, decorrelating angular positions breaks down the two-halo term, so that halos overlap less on the field of view and decreases high-peak counts. \cite{Yang_etal_2011} shows that high peaks with $\nu\geq4.8$ are mainly contributed by one single halo, and about 12\% of total high-peak counts are contributed by multiple halos. This number agrees with the undercount from our hypothesis of randomization. Combining this step with the former one, we confirm that considering lensing contribution from spatially decorrelated clusters is a good approximation for peak counts.

The impact of the mass function is shown by comparing Case 3 to Case 4. Peak counts are more numerous in our forward model based on the mass function of \cite{Jenkins_etal_2001}. This excess compensates for the deficit from randomization. However, as shown by \fig{fig:massComp}, the real mass function in $N$-body runs is coherent to the analytical model that we use, except for the low-mass deficit tails from $N$-body runs. To test the impact from this, we ran fast simulations with different lower limit for the halo sampling, and we discover that peak counts do not depend on the lower sampling limit $M_\min$ when $M_\min$ remains lower than $10^{13}\ \Msol/h$. This proves that the deficit tails are not the cause of the peak count enhancement. Without this explanation, we may have to test with another $N$-body simulation set to understand the origin of this effect.

\begin{figure}[htb]
	\centering
	\includegraphics[width=\columnwidth]{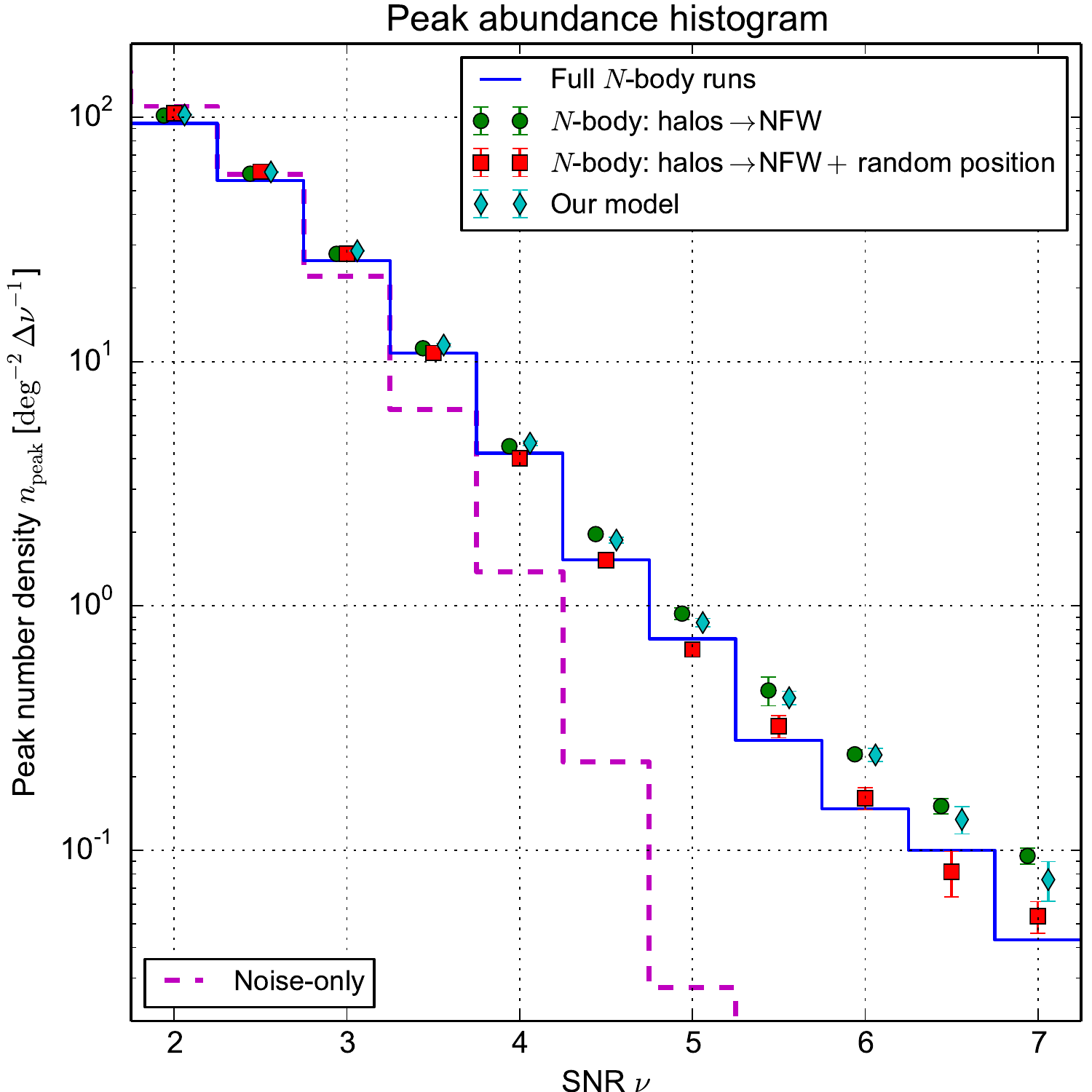}
	\caption{Similar plot to \fig{fig:small_field}, but in a larger field. Cases 2, 3, and 4 are carried out for 859~deg$^2$. Case 1 should only be taken as an indication, since its size of field is the same as in \fig{fig:small_field}, and therefore 16 times smaller than cases 2--4. The fluctuation from high $\nu$ bins is much reduced compared to \fig{fig:small_field}.}
	\label{fig:large_field}
\end{figure}

\figFull{fig:large_field} shows a similar study of Cases 2, 3, and 4 for a larger field of 859~deg$^2$. One can recover the same effects: compensation of effects deriving from NFW profiles and randomization. Therefore, the difference between our model and $N$-body simulations is at the same order of magnitude of the one between the analytical and the $N$-body mass functions. We would like to point out that the Poisson fluctuation has been largely suppressed. A quick calculation shows that, for a given peak density $n$ and a survey area $A$, the ratio of the Poisson noise to peak density is $1/\sqrt{nA}$. The error bars for high peaks in both \fig{fig:small_field} and \fig{fig:large_field} stay within 50\% of the values given by this formula. As a result, we argue that to reduce the Poisson fluctuation at the level of 10\%, a survey of more than 150 deg$^2$ is preferable using WL peaks with $\nu \lesssim 5.25$ and 800 deg$^2$ using peaks with $\nu \lesssim 6.25$.

\begin{figure}[htb]
	\centering
	\includegraphics[width=\columnwidth]{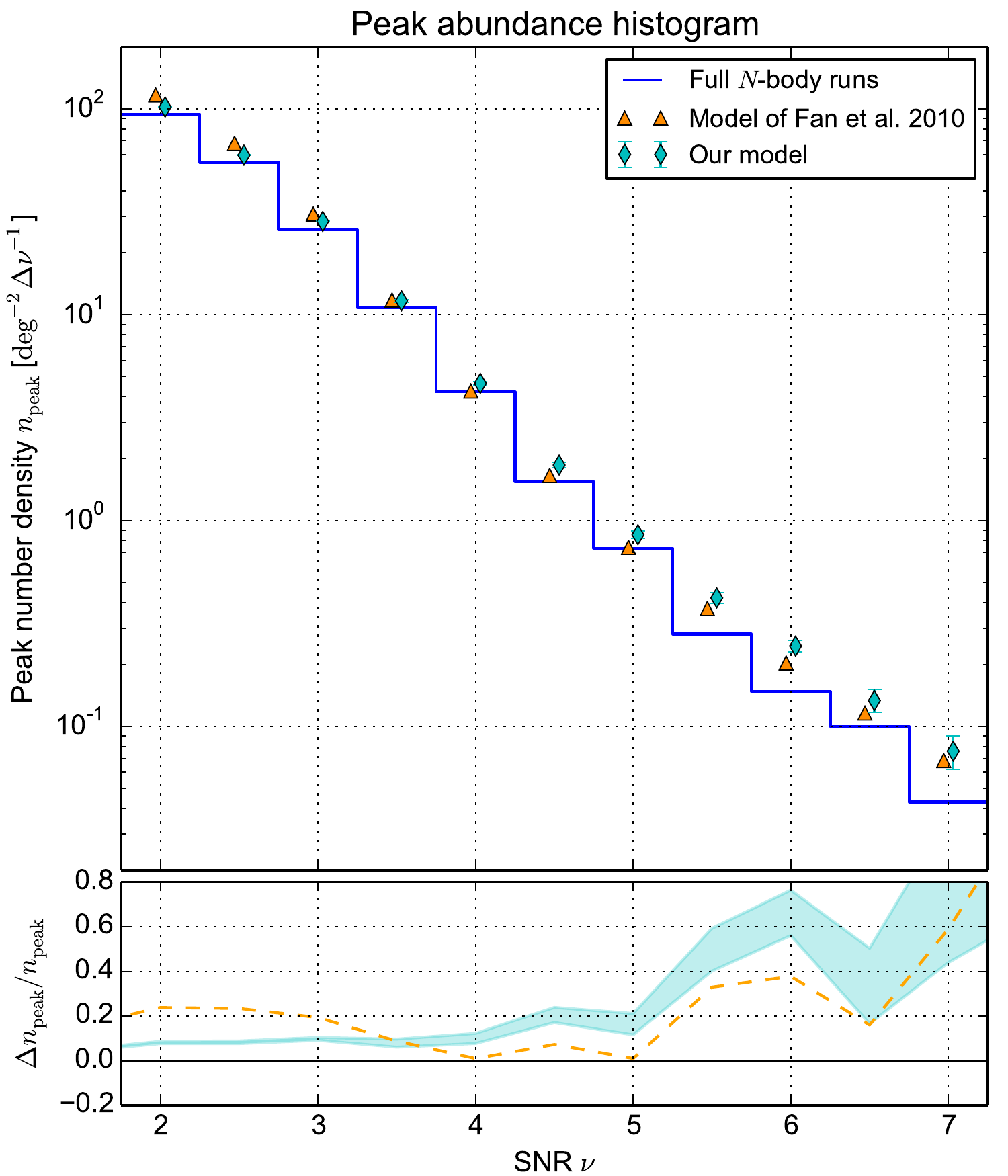}
	\caption{Comparison of the FSL model (orange triangles) to our model (cyan diamonds). The full $N$-body peak histogram is shown as a blue line. In the lower panel, we draw the difference between the FSL model and $N$-body data within an orange dashed line. The cyan-colored zone represents the error bars for our model. The field of view for fast simulations is 859 deg$^2$. The $N$-body data is only indicative.}
	\label{fig:Fan_vs_ours}
\end{figure}

\subsection{Comparison to an analytical model}
\label{subsec:Fan}

In \fig{fig:Fan_vs_ours}, we draw peak histograms obtained from the analytical model of \citetalias{Fan_etal_2010} and from our model. The computation for the \citetalias{Fan_etal_2010} model is done with the same halo profiles and parameters, and the same mass function. For our model, we use our large-field result as mentioned in the previous section. Both models are computed with the same parameter set as the Aardvark $N$-body simulation inputs. We observe that the \citetalias{Fan_etal_2010} model is also in good agreement with $N$-body runs. The prediction from the \citetalias{Fan_etal_2010} model is more consistent with $N$-body values for high-peak counts, whereas our model performs better in the low-peak regime. In general, the deviation of both models for $\nu\leq5.25$ stays under 25\%.

\subsection{Sensitivity tests on cosmological parameters}
\label{subsec:param}

\begin{figure*}[htb]
	\centering
	\includegraphics[width=17cm]{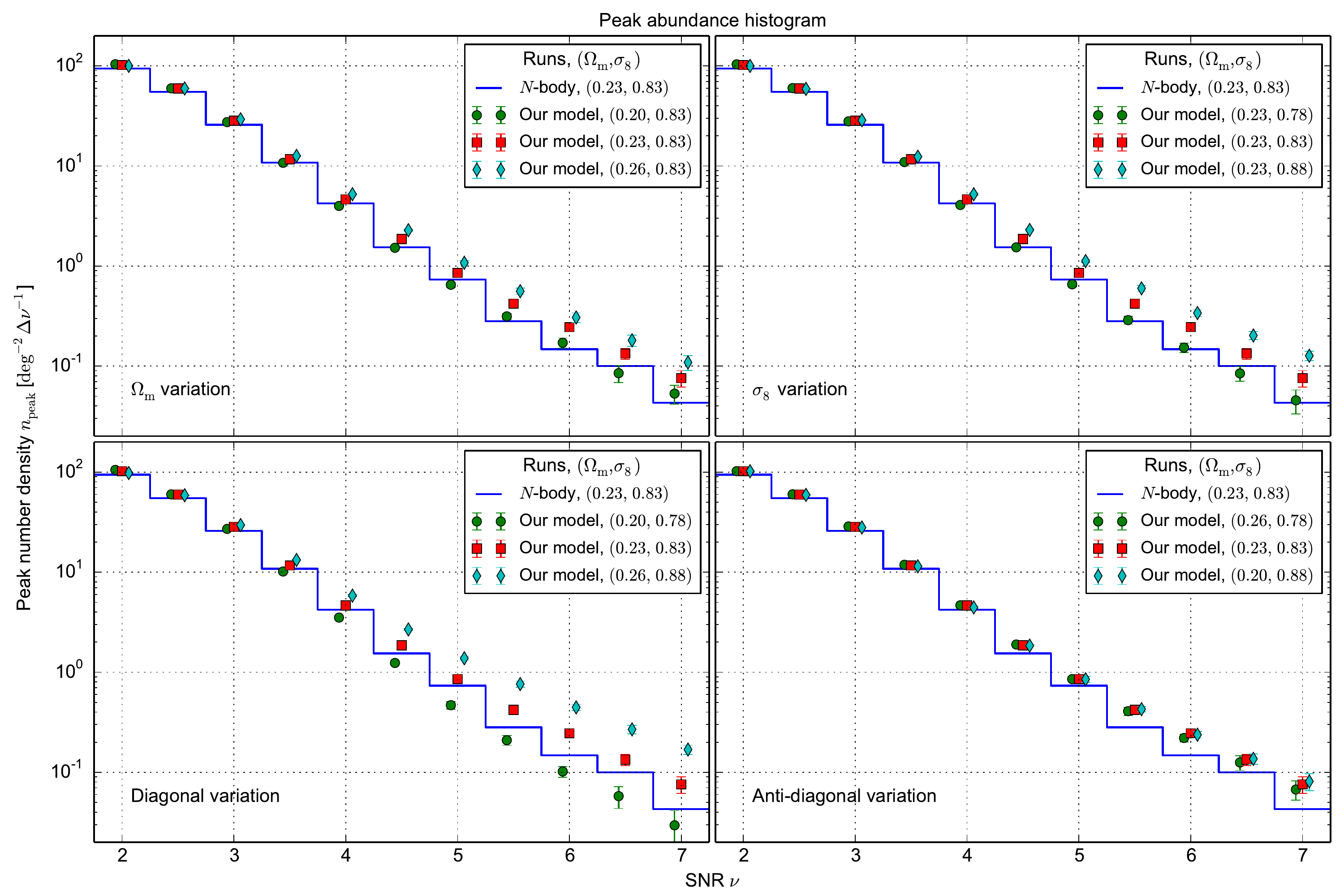}
	\caption{Sensitivity tests on $(\Omega_\mmmm,\sigma_8)$. The four plots indicate different variation directions in the $\Omega_\mmmm$-$\sigma_8$ plane. The upper left, upper right, lower left, and lower right panels show the variation in $\Omega_\mmmm$, $\sigma_8$, diagonal, and anti-diagonal direction, respectively. Blue solid lines represent full $N$-body runs, while red squares are always our model with $N$-body input. The field of view is 859 deg$^2$. The $N$-body data is only indicative.}
	\label{fig:param}
\end{figure*}

Finally in this section, we show how our model depends on cosmological parameters. Weak lensing is particularly sensitive to $\Omega_\mmmm$ and $\sigma_8$, hence we carry out nine series of fast simulations for which $(\Omega_\mmmm, \sigma_8)$ is chosen from $\left\{\Omega_\mmmm^{(N)}, \Omega_\mmmm^{(N)}\pm \Delta\Omega_\mmmm\right\}$ $\times$ $\left\{\sigma_8^{(N)}, \sigma_8^{(N)}\pm \Delta\sigma_8\right\}$, where $\Omega_\mmmm^{(N)}$ and $\sigma_8^{(N)}$ are input from our $N$-body runs. The values of $\Delta\Omega_\mmmm$ and $\Delta\sigma_8$ are chosen to be 0.03 and 0.05, respectively, and the remaining parameters are identical to the $N$-body simulations. Each scenario is the average over 16 combinations of four fast simulation realizations and four noise maps.

\figFull{fig:param} shows four plots that correspond to four variation directions on the $\Omega_\mmmm$-$\sigma_8$ parameter plane, with regard to $(\Omega_\mmmm^{(N)}, \sigma_8^{(N)})$. Both upper panels show the variation of only one parameter. They reveal that our model performs a neat, progressive difference of peak abundance in every bin, ranging from $\nu = 4$ to 6. We notice that the differences between cyan diamonds (higher value of $\Omega_\mmmm$ or $\sigma_8$) and red squares ($N$-body value) are always narrower than those between green circles (lower value of $\Omega_\mmmm$ or $\sigma_8$) and red squares. This is triggered by the banana-shape constraint on the $\Omega_\mmmm$-$\sigma_8$ plane, from which a horizontal or a vertical cut will result in an asymmetric confidence level for a single parameter.

The two lower panels are variations in the diagonal and anti-diagonal directions. Like what we expect, the diagonal variation is the most efficient discriminant of $\Omega_\mmmm$-$\sigma_8$. In contrast, peak counts for different parameter sets completely merge together in the lower right-hand panel, since the anti-diagonal direction corresponds roughly to the degenerency lines. Furthermore, all error bars (for $3.75 \leq \nu \leq 6.25$) remain smaller than 5\%, which shows the robustness of our model. We recall that blue solid lines correspond to a small 53.7 deg$^2$ field, such that the Poisson noise might bias high peak counts, as explained in \sect{subsec:validation}. At the end of the day, the performance of our model at distinguishing different cosmological models has been confirmed.

\figFull{fig:param} also shows that systematic biases of our model could lead to parameter biases. A simple interpolation for the bin $\nu=5$ shows that $N$-body peak counts correspond to a cosmology with $\Omega_\mmmm \approx 0.212$ if the knowledge of $\sigma_8$ is perfect. The bias is then $\Delta\Omega_\mmmm \approx 0.018$. Similarly, the bias on $\sigma_8$ is $\Delta\sigma_8 \approx 0.030$ if $\Omega_\mmmm$ is known. The origin of the biases of our model is complex. We discuss a list of possible improvements that reduce potential systematics in the following section.

\section{Summary and perspectives}
\label{sec:conclu}

WL peaks probe cosmological structures in a straightforward way, since they are directly induced by total-mass gravitational effects, and they especially probe the high-mass part of the mass function. Unlike other tracers, WL peaks provide a forward-fitting approach to study the mass function and cosmology. This makes WL peaks a very competitive candidate for improving our knowledge about structure formation.

In this paper, we presented a new model that predicts weak-lensing peak counts. We generated fast simulations by sampling halos from analytical expressions. By assuming that halos in these simulations are randomly distributed on the sky, we count peaks from ray-tracing maps obtained from these simulations to predict number counts. In this model, we have supposed that unbound matter contributes little to the lensing and that halo clustering has little impact on peak counts.

We validated our approach by comparing number counts with $N$-body results. In particular, we focused on peaks with $\nu \approx$~4--6, since lower $\nu$ are dominated by shape noise, and higher $\nu$ are dominated by the Poisson fluctuation. We showed how the three steps corresponding to the main assumptions of our model influence convergence peak abundance. 
First, NFW profiles tend to shift some medium peaks to higher values, in spite of the lack of unbound objects. Second, the number of peaks decreases when halo positions are randomized. Last, the difference between the $N$-body mass function and the analytical one is observable in produced peak counts. In summary, our model is in good agreement with results from full $N$-body runs.

We also tested the dependence of our model on $\Omega_\mmmm$ and $\sigma_8$. For a 859 deg$^2$ sky area, the Poisson fluctuation is reduced to a reasonable level for peaks with $\nu\lesssim 6$. It turns out that different scenarios are discernable for $\nu \gtrsim 4$, with a degerency direction corresponding roughly to the anti-diagonal in the plane $\Omega_\mmmm$-$\sigma_8$. Tests on a large set of different parameters are feasible with our model thanks to the short computation time.

Our probabilistic model has other potential advantages. Repeated simulations for the same cosmological parameters generate the distribution of observables. This allows us to compare observations with our model without the need to define a likelihood function or to assume any Gaussian distribution. For example, model discrimination can be carried out using the false discovery rate method \citep[FDR, ][an application can be found in \citealt{Pires_etal_2009a}]{Benjamini_Hochberg_1995}, approximate Bayesian computation \citep[ABC, see for example][]{Cameron_Pettitt_2012, Weyant_etal_2013}, or other statistical techniques. Another powerful advantage of our model is its flexibility. Additional effects such as intrinsic ellipticity alignment, alternative methods such as nonlinear filters, and realistic survey settings, such as mask effects, magnification bias \citep{Liu_etal_2014a}, shape measurement errors \citep{Bard_etal_2013}, and photo-$z$ errors, can all be modeled in this peak-counting framework. The forward-modeling approach allows for a straightforward inclusion and marginalization of model uncertainties and systematics.

Several improvements to our model are possible. Using perturbation theory, we may take halo clustering into account in fast simulations. This can be done by some fast algorithms, such as \textsc{PTHalos} \citep{Scoccimarro_Sheth_2002}, \textsc{Pinocchio} \citep[][see also \citealt{Heisenberg_etal_2011}]{Monaco_etal_2002}, and remapping LPT \citep{Leclercq_etal_2013}. In addition, we can go beyond the idealized setting considered in this work by including a realistic source distribution, intrinsic alignment, mask effects, etc. We also expect that nonlinear filters and tomography studies may bring some more refined results for cosmology from peak counting. Finally, peak counts can be supplemented with additional WL observables, such as magnification and flexion.

The \textsc{Camelus} algorithm is implemented in C language. It requires the \textsc{Nicaea} library for cosmological computations. The \textsc{Camelus} source code is released via the website \footnote{\url{http://www.cosmostat.org/software/camelus/}}.

\begin{acknowledgements}
	This study is supported by R\'egion d'\^Ile-de-France under grant DIM-ACAV and the French national program for cosmology and galaxies (PNCG). The authors acknowledge the anonymous referee for useful comments and suggestions. We would like to thank August Evrard for providing $N$-body simulations. We also thank Zuhui Fan, Xiangkun Liu, and Chuzhong Pan for giving constructive comments on the preprint. Chieh-An Lin is very grateful for inspiring discussions with Fran\c{c}ois Lanusse, Yohan Dubois, and Michael Vespe. 
\end{acknowledgements}

\bibliography{Bibliographie_Linc}

\begin{thebibliography}{44}
\expandafter\ifx\csname natexlab\endcsname\relax\def\natexlab#1{#1}\fi

\bibitem[{{Bard} {et~al.}(2013){Bard}, {Kratochvil}, {Chang}, {May}, {Kahn},
  {AlSayyad}, {Ahmad}, {Bankert}, {Connolly}, {Gibson}, {Gilmore}, {Grace},
  {Haiman}, {Hannel}, {Huffenberger}, {Jernigan}, {Jones}, {Krughoff},
  {Lorenz}, {Marshall}, {Meert}, {Nagarajan}, {Peng}, {Peterson}, {Rasmussen},
  {Shmakova}, {Sylvestre}, {Todd}, \& {Young}}]{Bard_etal_2013}
{Bard}, D., {Kratochvil}, J.~M., {Chang}, C., {et~al.} 2013, \apj, 774, 49

\bibitem[{Bardeen {et~al.}(1986)Bardeen, Bond, Kaiser, \&
  Szalay}]{Bardeen_etal_1986}
Bardeen, J.~M., Bond, J.~R., Kaiser, N., \& Szalay, A.~S. 1986, \apj, 304, 15

\bibitem[{Bartelmann(1996)}]{Bartelmann_1996}
Bartelmann, M. 1996, \aap, 313, 697

\bibitem[{Becker(2013)}]{Becker_2013}
Becker, M.~R. 2013, \mnras, 435, 115

\bibitem[{Behroozi {et~al.}(2013)Behroozi, Wechsler, \&
  Wu}]{Behroozi_etal_2013}
Behroozi, P.~S., Wechsler, R.~H., \& Wu, H.-Y. 2013, \apj, 762, 109

\bibitem[{Benjamini \& Hochberg(1995)}]{Benjamini_Hochberg_1995}
Benjamini, Y. \& Hochberg, Y. 1995, Journal of the Royal Statistical Society,
  Series B, 57, 289–300

\bibitem[{Bond \& Efstathiou(1987)}]{Bond_Efstathiou_1987}
Bond, J.~R. \& Efstathiou, G. 1987, \mnras, 226, 655

\bibitem[{{Bullock} {et~al.}(2001){Bullock}, {Kolatt}, {Sigad}, {Somerville},
  {Kravtsov}, {Klypin}, {Primack}, \& {Dekel}}]{Bullock_etal_2001}
{Bullock}, J.~S., {Kolatt}, T.~S., {Sigad}, Y., {et~al.} 2001, \mnras, 321, 559

\bibitem[{{Cameron} \& {Pettitt}(2012)}]{Cameron_Pettitt_2012}
{Cameron}, E. \& {Pettitt}, A.~N. 2012, \mnras, 425, 44

\bibitem[{{Clerc} {et~al.}(2012){Clerc}, {Pierre}, {Pacaud}, \&
  {Sadibekova}}]{Clerc_etal_2012}
{Clerc}, N., {Pierre}, M., {Pacaud}, F., \& {Sadibekova}, T. 2012, \mnras, 423,
  3545

\bibitem[{Dietrich \& Hartlap(2010)}]{Dietrich_Hartlap_2010}
Dietrich, J.~P. \& Hartlap, J. 2010, \mnras, 402, 1049

\bibitem[{Fan {et~al.}(2010)Fan, Shan, \& Liu}]{Fan_etal_2010}
Fan, Z., Shan, H., \& Liu, J. 2010, \apj, 719, 1408

\bibitem[{G{\'o}rski {et~al.}(2005)G{\'o}rski, Hivon, Banday, Wandelt, Hansen,
  Reinecke, \& Bartelmann}]{Gorski_etal_2005}
G{\'o}rski, K.~M., Hivon, E., Banday, A.~J., {et~al.} 2005, \apj, 622, 759

\bibitem[{Hamana {et~al.}(2004)Hamana, Takada, \& Yoshida}]{Hamana_etal_2004}
Hamana, T., Takada, M., \& Yoshida, N. 2004, \mnras, 350, 893

\bibitem[{{Heisenberg} {et~al.}(2011){Heisenberg}, {Sch{\"a}fer}, \&
  {Bartelmann}}]{Heisenberg_etal_2011}
{Heisenberg}, L., {Sch{\"a}fer}, B.~M., \& {Bartelmann}, M. 2011, \mnras, 416,
  3057

\bibitem[{Hennawi \& Spergel(2005)}]{Hennawi_Spergel_2005}
Hennawi, J.~F. \& Spergel, D.~N. 2005, \apj, 624, 59

\bibitem[{{Jain} \& {Van Waerbeke}(2000)}]{Jain_VanWaerbeke_2000}
{Jain}, B. \& {Van Waerbeke}, L. 2000, \apjl, 530, L1

\bibitem[{Jenkins {et~al.}(2001)Jenkins, Frenk, White, Colberg, Cole, Evrard,
  Couchman, \& Yoshida}]{Jenkins_etal_2001}
Jenkins, A., Frenk, C.~S., White, S.~D.~M., {et~al.} 2001, \mnras, 321, 372

\bibitem[{{Kainulainen} \& {Marra}(2009)}]{Kainulainen_Marra_2009}
{Kainulainen}, K. \& {Marra}, V. 2009, \prd, 80, 123020

\bibitem[{{Kainulainen} \&
  {Marra}(2011{\natexlab{a}})}]{Kainulainen_Marra_2011}
{Kainulainen}, K. \& {Marra}, V. 2011{\natexlab{a}}, \prd, 83, 023009

\bibitem[{{Kainulainen} \&
  {Marra}(2011{\natexlab{b}})}]{Kainulainen_Marra_2011a}
{Kainulainen}, K. \& {Marra}, V. 2011{\natexlab{b}}, \prd, 84, 063004

\bibitem[{{Kratochvil} {et~al.}(2010){Kratochvil}, {Haiman}, \&
  {May}}]{Kratochvil_etal_2010}
{Kratochvil}, J.~M., {Haiman}, Z., \& {May}, M. 2010, \prd, 81, 043519

\bibitem[{{Leclercq} {et~al.}(2013){Leclercq}, {Jasche}, {Gil-Mar{\'{\i}}n}, \&
  {Wandelt}}]{Leclercq_etal_2013}
{Leclercq}, F., {Jasche}, J., {Gil-Mar{\'{\i}}n}, H., \& {Wandelt}, B. 2013,
  \jcap, 11, 48

\bibitem[{{Liu} {et~al.}(2014){Liu}, {Haiman}, {Hui}, {Kratochvil}, \&
  {May}}]{Liu_etal_2014a}
{Liu}, J., {Haiman}, Z., {Hui}, L., {Kratochvil}, J.~M., \& {May}, M. 2014,
  \prd, 89, 023515

\bibitem[{{Mainini} \& {Romano}(2014)}]{Mainini_Romano_2014}
{Mainini}, R. \& {Romano}, A. 2014, \jcap, 8, 63

\bibitem[{{Marian} {et~al.}(2010){Marian}, {Smith}, \&
  {Bernstein}}]{Marian_etal_2010}
{Marian}, L., {Smith}, R.~E., \& {Bernstein}, G.~M. 2010, \apj, 709, 286

\bibitem[{Maturi {et~al.}(2010)Maturi, Angrick, Pace, \&
  Bartelmann}]{Maturi_etal_2010}
Maturi, M., Angrick, C., Pace, F., \& Bartelmann, M. 2010, \aap, 519, A23

\bibitem[{{Monaco} {et~al.}(2002){Monaco}, {Theuns}, \&
  {Taffoni}}]{Monaco_etal_2002}
{Monaco}, P., {Theuns}, T., \& {Taffoni}, G. 2002, \mnras, 331, 587

\bibitem[{Navarro {et~al.}(1996)Navarro, Frenk, \& White}]{Navarro_etal_1996}
Navarro, J.~F., Frenk, C.~S., \& White, S. D.~M. 1996, \apj, 462, 563

\bibitem[{Navarro {et~al.}(1997)Navarro, Frenk, \& White}]{Navarro_etal_1997}
Navarro, J.~F., Frenk, C.~S., \& White, S. D.~M. 1997, \apj, 490, 493

\bibitem[{Pires {et~al.}(2012)Pires, Leonard, \& Starck}]{Pires_etal_2012}
Pires, S., Leonard, A., \& Starck, J.-L. 2012, \mnras, 423, 983

\bibitem[{{Pires} {et~al.}(2009){Pires}, {Starck}, {Amara},
  {R{\'e}fr{\'e}gier}, \& {Teyssier}}]{Pires_etal_2009a}
{Pires}, S., {Starck}, J.-L., {Amara}, A., {R{\'e}fr{\'e}gier}, A., \&
  {Teyssier}, R. 2009, \aap, 505, 969

\bibitem[{Schneider {et~al.}(1998)Schneider, {Van Waerbeke}, Jain, \&
  Kruse}]{Schneider_etal_1998}
Schneider, P., {Van Waerbeke}, L., Jain, B., \& Kruse, G. 1998, \mnras, 296,
  873

\bibitem[{{Scoccimarro} \& {Sheth}(2002)}]{Scoccimarro_Sheth_2002}
{Scoccimarro}, R. \& {Sheth}, R.~K. 2002, \mnras, 329, 629

\bibitem[{{Shan} {et~al.}(2014){Shan}, {Kneib}, {Comparat}, {Jullo},
  {Charbonnier}, {Erben}, {Makler}, {Moraes}, {Van Waerbeke}, {Courbin},
  {Meylan}, {Tao}, \& {Taylor}}]{Shan_etal_2014}
{Shan}, H.~Y., {Kneib}, J.-P., {Comparat}, J., {et~al.} 2014, \mnras, 442, 2534

\bibitem[{Springel(2005)}]{Springel_2005}
Springel, V. 2005, \mnras, 364, 1105

\bibitem[{Takada \& Jain(2002)}]{Takada_Jain_2002}
Takada, M. \& Jain, B. 2002, \mnras, 337, 875

\bibitem[{Takada \& Jain(2003)}]{Takada_Jain_2003a}
Takada, M. \& Jain, B. 2003, \mnras, 340, 580

\bibitem[{{Van Waerbeke}(2000)}]{VanWaerbeke_2000}
{Van Waerbeke}, L. 2000, \mnras, 313, 524

\bibitem[{Wang {et~al.}(2009)Wang, Haiman, \& May}]{Wang_etal_2009}
Wang, S., Haiman, Z., \& May, M. 2009, \apj, 691, 547

\bibitem[{{Weinberg} \& {Kamionkowski}(2003)}]{Weinberg_Kamionkowski_2003}
{Weinberg}, N.~N. \& {Kamionkowski}, M. 2003, \mnras, 341, 251

\bibitem[{{Weyant} {et~al.}(2013){Weyant}, {Schafer}, \&
  {Wood-Vasey}}]{Weyant_etal_2013}
{Weyant}, A., {Schafer}, C., \& {Wood-Vasey}, W.~M. 2013, \apj, 764, 116

\bibitem[{{Yang} {et~al.}(2013){Yang}, {Kratochvil}, {Huffenberger}, {Haiman},
  \& {May}}]{Yang_etal_2013}
{Yang}, X., {Kratochvil}, J.~M., {Huffenberger}, K., {Haiman}, Z., \& {May}, M.
  2013, \prd, 87, 023511

\bibitem[{Yang {et~al.}(2011)Yang, Kratochvil, Wang, Lim, Haiman, \&
  May}]{Yang_etal_2011}
Yang, X., Kratochvil, J.~M., Wang, S., {et~al.} 2011, \prd, 84, 043529

\end{thebibliography}

\end{document}